\documentclass{nature}
\usepackage{amsmath}    
\usepackage{graphicx}   
\usepackage{color}
\usepackage[10pt]{moresize}
\usepackage{graphics,graphicx}
\usepackage{amsfonts}
\usepackage{amssymb}
\usepackage{amscd}
\usepackage{amsmath}
\usepackage{enumerate}
\usepackage{epsfig}
\usepackage{subfigure}
\usepackage{epstopdf}
\usepackage{bm}

\begin{document}

\title{Gain enhanced Fano resonance in a coupled photonic crystal cavity-waveguide structure}
\author{Yanhui Zhao, Chenjiang Qian, Kangsheng Qiu, Jing Tang, Yue Sun, Kuijuan Jin, Xiulai Xu$^{\dag}$}
\maketitle
\begin{affiliations}
\item
Beijing National Laboratory for Condensed Matter Physics, Institute
of Physics, Chinese Academy of
Sciences, Beijing 100190, P.~R.~China \\
$^\dag$ To whom correspondence should be addressed. E-mail:
xlxu@iphy.ac.cn
\end{affiliations}

\baselineskip24pt

\maketitle

\begin{abstract}
Systems with coupled cavities and waveguides have been demonstrated as optical switches and optical sensors. To optimize the functionalities of these optical devices, Fano resonance with asymmetric and steep spectral line shape has been used. We theoretically propose a coupled photonic crystal cavity-waveguide structure to achieve Fano resonance by placing partially reflecting elements in waveguide. To enhance Fano resonance, optical gain material is introduced into the cavity. As the gain increases, the transmission line shape becomes steepened and the transmissivity can be six times enhanced, giving a large contrast by a small frequency shift. It is prospected that the gain enhanced Fano resonance is very useful for optical switches and optical sensors.
\end{abstract}

\section*{Introduction}

Photonic crystal (PhC) cavities \cite{PhysRevE.65.016608,akahane2003high} have high quality factors and small mode volume. With such cavities, light can be confined in a very small volume with low loss rate. Thus, the interaction between light and matter can be greatly enhanced, which is important for investigations of cavity quantum electrodynamics~\cite{vuvckovic2003photonic,PhysRevE.65.016608,brossard2010strongly,tang2015quantum}. PhC waveguide can guide light with very low losses \cite{jamois2003silicon, watanabe2007topology}. The coupled PhC cavity-waveguide structures have been used to realize various types of optical devices, such as optical switches \cite{Heuck:13,Nozaki:13,belotti2008all,nozaki2010sub,tanabe2005all,:/content/aip/journal/apl/105/6/10.1063/1.4893451,Zhao:15} and optical sensors \cite{yang2011nanoscale,shanthi2014two}. To improve the device performance, Fano resonance with asymmetric and steep spectral line shape has been intensively used, which arises from the interference between discrete resonance states and continuum states~\cite{PhysRev.124.1866}. By modulating the coupling between PhC cavity and waveguide, Fano resonance can be achieved due to the constructive and destructive interference of discrete cavity resonance states with broadband continuum waveguide states \cite{:/content/aip/journal/apl/105/6/10.1063/1.4893451,Heuck:13,Nozaki:13}.

In this paper,  we design a coupled PhC cavity \cite{:/content/aip/journal/apl/83/21/10.1063/1.1629140} and waveguide structure using Finite-difference time-domain (FDTD) simulation \cite{taflove1995computational}. By placing partially reflecting elements in the PhC waveguide, Fano type transmission line shape is achieved. Recently, gain materials such as quantum dots~\cite{xu2007,xu2008}, nanocrystals \cite{sun2016} and rare-earth ions \cite{Chang_2014,Peng_2014} have been used as light-emitting centers to achieve single-photon sources, display devices and all-optical diodes, respectively. Additionally, lasers have been demonstrated with coupling these light sources to microcavities \cite{song2010,Nomura:09}. Here, such gain materials are introduced into PhC cavities to enhance Fano resonance. With introducing the gain materials into cavity, the transmission line shape of the Fano resonance is greatly steepened and a large contrast is achieved by a small frequency shift. In our structure, the transmissivity in the waveguide is increased by a factor of six, which is potentially applicable for optical switches and optical sensors.

\section*{Results}
\subsection{Coupled PhC H1 cavity and waveguide structure.}
We employed a free-software package MEEP \cite{oskooi2010meep} to design the PhC structure. Hexagonal lattice of air holes was patterned on a planar dielectric slab. The slab thickness is 0.6 $\textit{a}$, where \textit{a} is the lattice constant. The air hole radius is $0.3 \textit{a}$, and the slab dielectric is 12.96. With the MIT Photonic-Bands (MPB) package \cite{Johnson2001:mpb}, the PhC slab band structure has been calculated, which has TE-like photonic bandgap from  0.2455 to 0.3229 (\textit{a}/$\lambda $), where $\lambda $ is wavelength in the vacuum. With one air hole missing in the PhC slab, an H1 cavity is formed \cite{shirane2007}. The radii of the nearest six air holes around the H1 cavity are reduced to 0.24 \textit{a} to increase the quality factors of the cavity modes. Three lattice constant \textit{a} away from the center of H1 cavity, one array of air holes are removed to form a W1 waveguide. \mbox{Figure 1(a)} shows the coupled PhC cavity-waveguide structure.

 To simulate the spectral response of the coupled cavity-waveguide structure, a Gaussian source with frequency center 0.2980 (\textit{a}/$ \lambda $) and frequency width 0.02 (\textit{a}/$ \lambda $) is located at the left end of waveguide and the energy flux has been detected at the right end. \mbox{Figure 1(b)} shows the normalized transmission spectrum of the light. As waveguide mode frequencies are far away from the cavity resonant frequencies, light transmits from one end of the waveguide to another end with almost 100\% transmission without any coupling with the cavity. When waveguide modes are resonant or nearly resonant with the cavity modes, light couples into the cavity resulting in dips in the transmission spectrum. Three dips in the transmission spectrum in Fig. 1(b) correspond to three resonant modes of the H1 cavity.

\mbox{Figure 2} shows the z components of the magnetic fields of the cavity modes corresponding to the three dips in the transmission spectrum in \mbox{Fig. 1(b)}. The mode of \mbox{Fig. 2(a)} shows a hexapole mode corresponding to the dip denoted by \textit{a}. While the modes of \mbox{Fig. 2(b)} and (c) are nearly degenerate quadrupole modes for \textit{b} and \textit{c} in \mbox{Fig. 1(b)}. The cavity mode field distributions indicate that the cavity couples with the waveguide via the evanescent field. The cavity resonance frequencies of \mbox{Fig. 2(a)}, (b) and (c) are 0.2961, 0.2996 and 0.3002 (\textit{a}/$ \lambda $) and the corresponding quality factors are 2890, 5960 and 2770, respectively. A cavity mode with low quality factor has a large coupling strength with the waveguide, thus a deeper dip appears in the transmission spectrum with a broad linewidth.

\subsection{The realization of Fano resonance via partially reflecting elements.}
To realize Fano resonance, partially reflecting elements are introduced into the coupled PhC cavity-waveguide structure. Two air hole barriers have been placed in the center of the waveguide, as shown in \mbox{Fig. 3(a)}. The amplitude transmissivity $t(\omega)$ is determined by \cite{yu2015ultrafast}
\begin{equation}
t(\omega)=\frac{t_{B}(\omega_{0}-\omega)\pm 2\gamma\sqrt{1-t_{B}^{2}}-it_{B}\gamma_{v}}{i(\omega_{0}-\omega)+\gamma_{v}+2\gamma}
\end{equation}
 where $\omega_{0}$ is the resonant frequency of the cavity mode. $t_{B}$ is the amplitude transmission coefficient of the partially reflecting elements. $\gamma_{v}$ is the intrinsic cavity loss rate. $\gamma$ is the cavity decay rate due to coupling into one port of the waveguide. When there is no partially reflecting elements existing ($t_{B}=1$), the transmission line shape is Lorentzian symmetric. When $t_{B}\neq1$, the light from the incoming port of waveguide to the outgoing port has two pathways. One with light transmitting through the cavity provides discrete modes, and the other with light passing through the partially reflecting elements gives rise to continuum modes. The constructive and destructive interferences between these modes induce an asymmetric Fano type transmission line shape. The amplitude transmission coefficient $t_{B}$ determines whether the transmission line shape is symmetric Lorentzian type or asymmetric Fano type.

The simulated and normalized transmission spectrum is shown in \mbox{Fig. 3(b)}. Two Fano type peaks are observed in the spectrum. When the waveguide modes are non-resonant with the cavity modes, the light can not be coupled into the cavity and is reflected mostly by the two air hole barriers. The light with the cavity mode frequency can couple into the cavity and transmit to the other port of the waveguide. Therefore, Fano type peaks appear in the transmission spectrum. The simulated cavity mode distributions show that, there are three cavity modes in the frequency range of the transmission spectrum in \mbox{Fig. 3(b)}. The z components of the magnetic fields of the three modes are shown in \mbox{Fig. 4(a)}, (b) and (c). The cavity resonance frequencies of the three modes are 0.2956, 0.2998 and 0.2998 (\textit{a}/$ \lambda $) and the corresponding quality factors are 1830, 2980 and 4570, respectively. The resonant frequencies of cavity modes in \mbox{Fig. 4(b)} and (c) are so close that the cavity modes can not be distinguished in the transmission spectrum in \mbox{Fig. 3(b)}.

\subsection{Gain enhanced Fano resonance.}
The asymmetric and steep line shape of Fano resonance is beneficial for optical switches and optical sensors. The steep line shape of Fano resonance can significantly reduce the frequency shift required for on/off switching. Optical sensors with high sensitivity can be constructed based on the spectral shift associated with asymmetric line shapes. Recently, investigations have been performed to enhance the Fano resonance. Heuck \textit{et al.} \cite{Heuck:13} reported that the Fano resonance can be modified by adjusting the partially reflecting elements. In this work, we introduce gain material into the PhC cavity to enhance the Fano resonance.

In the two dimensional PhC slab structure, it is straightforward to couple quantum dots into a cavity \cite{brossard2010strongly,tang2015quantum}, with which the optical gain can be obtained by pumping the quantum dot either optically or electrically. As the quantum dot transition and the cavity mode are tuned into resonance, the loss of the cavity mode is compensated by the photons emitted by the quantum dot, resulting in increasing of the quality factor of the cavity mode. To simulate the gain supplied by quantum dot, we consider the material structure in Maxwell's equations which is determined by the relative permittivity $\varepsilon(\bm{x})$. It should be noted that $\varepsilon$ does not depend only on position, but also on frequency (material dispersion), electric field (nonlinearity), and the orientation of the field (anisotropy) \cite{oskooi2010meep}. Material dispersion is generally associated with absorption loss or gain in the material, which is given by the following expression:
\begin{equation}
\begin{split}
\varepsilon\left(f,\bm{x}\right) = &\left(1+\dfrac{\textit{i}\cdot\sigma_{D}\left( \bm{x}\right) }{2\pi f} \right) \\
&\cdot\left[ \varepsilon_{\infty}\left( \bm{x}\right)+ \sum_{n}\dfrac{\sigma_{n}\left( \bm{x}\right)\cdot f_{n}^{2} }{f_{n}^{2}-f^{2}-\textit{i}f\gamma_{n}/2\pi}  \right]
\end{split}
\end{equation}
where $\varepsilon_{\infty}$ is the instantaneous dielectric function, $\sigma_{D}$ is the electric conductivity, $f_{n}$ and $\gamma_{n}$ are user-specified constants, and $\sigma_{n}\left(\bm{x} \right)$ is a user-specified function of position giving the strength of the \textit{n}-th resonance. Note that the imaginary part of $\varepsilon$ can be expressed by conductivity $\sigma_{D}$ as $\textit{i}\varepsilon_{\infty}\sigma_{D}/2\pi f$. Here, we only consider the absorption loss or gain in a narrow bandwidth, thus the imaginary part of $\varepsilon$ can be set to some known experimental value, and a dispersionless real part of $\varepsilon$ can be set with a value in our bandwidth of interest. In FDTD simulation, the imaginary part Im$\varepsilon=\varepsilon_{\infty}\sigma_{D}/2\pi f$ is set by choosing the conductivity $\sigma_{D}$ at our frequency $f$ of interest. The absorption loss (gain) in the material is associated with positive (negative) value of the imaginary part of $\varepsilon$. A series of negative values of Im$\varepsilon$ has been used to simulate the gain, which can be achieved by pumping quantum dot in the cavity.

In our coupled PhC cavity-waveguide structure (shown in \mbox{Fig. 3(a)}), gain material is located at the center of the H1 cavity with a radius of 0.6 $a$ and thickness of 0.6 $a$. The frequency $f$ is set around the resonant frequency of the degenerated quadrupole modes. The imaginary part of the dielectric function of the gain material introduced into the H1 cavity is set as -0.001, -0.01, -0.0135 and -0.01415 at frequency 0.2998 $(a/\lambda)$. The normalized transmission spectra are shown in \mbox{Fig. 5}. As the gain increases, the normalized transmissivity becomes larger and larger, and the linewidth of the transmission line shape becomes narrower. Light trapped in the cavity is fed by the gain so that the light energy flux is enlarged. Then the enlarged light energy flux is coupled into waveguide, resulting in enlarged transmissivity. Without gain material in the cavity, the normalized transmissivity corresponding to the degenerated quadrupole modes is 0.246. When gain material with Im$\varepsilon$=0.01415 is introduced into the cavity, the normalized transmissivity is 1.564, which is about six times larger than that of the case without gain. It can be clearly seen that the line shapes become steeper as the gain increases.

With the coupled cavity-waveguide structure, we further analysed the transmission spectra in the cases with and without gain material in H1 cavity. With the same frequency shift, the transmission contrast with gain material is much larger than the case without the gain. The large transmission contrast with a small frequency shift is very promising for realizing optical switches \cite{Heuck:13,Nozaki:13}. In addition, the resonant mode can be easily perturbed if there are nano-particles adhere to the H1 cavity, which gives a frequency shift in the transmission spectrum. The narrow and steep Fano asymmetric transmission lineshape could make the coupled structure much more attractive as a nano-particle sensor \cite{Shao2013}.

\section*{Discussion}
We have theoretically proposed that the Fano resonance in a coupled PhC cavity-wavegudie structure can be greatly enhanced when gain material is introduced into the cavity. The Fano spectral line shape becomes steeper and steeper as the gain increases. With Im$\varepsilon=-0.01415$ the transmission is enhanced by six times in our case. The optical gain in the cavity supplies photons with frequency of the cavity mode. The supplied photons compensate the losses of cavity mode. As a result, the quality factor of the cavity mode is increased, corresponding to a narrower linewidth. The discrete cavity mode with narrower linewidth interferes with the continuum modes, generating a steeper Fano asymmetric linedwidth. The doubly degenerate modes at 0.2998 $(a/\lambda)$ are chosen to enhance Fano asymmetric lineshape because the field of the degenerate modes distributes in a large volume in the cavity, which can enhance the gain. With the gain enhanced Fano resonance, a large contrast can be achieved by only a small frequency shift, which could be useful in ultralow-energy and high-contrast optical switches. The steepened transmission line shape can be also sensitive to nano-particle adhering to the designed structures, which enables such coupled PhC structure to be used as optical sensors.

\noindent\textbf{Acknowledgments}\\
This work was supported by the National Basic
Research Program of China under Grant No.
2013CB328706 and 2014CB921003; the National
Natural Science Foundation of China under
Grant No. 91436101, 11174356 and 61275060; the
Strategic Priority Research Program of the
Chinese Academy of Sciences under Grant No.
XDB07030200; and the Hundred Talents Program
of the Chinese Academy of Sciences.

\noindent\textbf{Author Contributions}\\
X.X. and K.J. conceived the project. Y.Z. designed the structure and carried out the calculations. Y.Z., C.Q., K.Q., J.T., Y.S. and X.X. discussed the results and the experimental implementations. Y.Z. and X.X. co-wrote the paper and all authors reviewed the manuscript.

\noindent\textbf{Additional Information}\\
The authors declare no competing financial interests.

\bibliographystyle{naturemag}


\clearpage
\begin{figure}
\centering \resizebox{16cm}{!}{\includegraphics{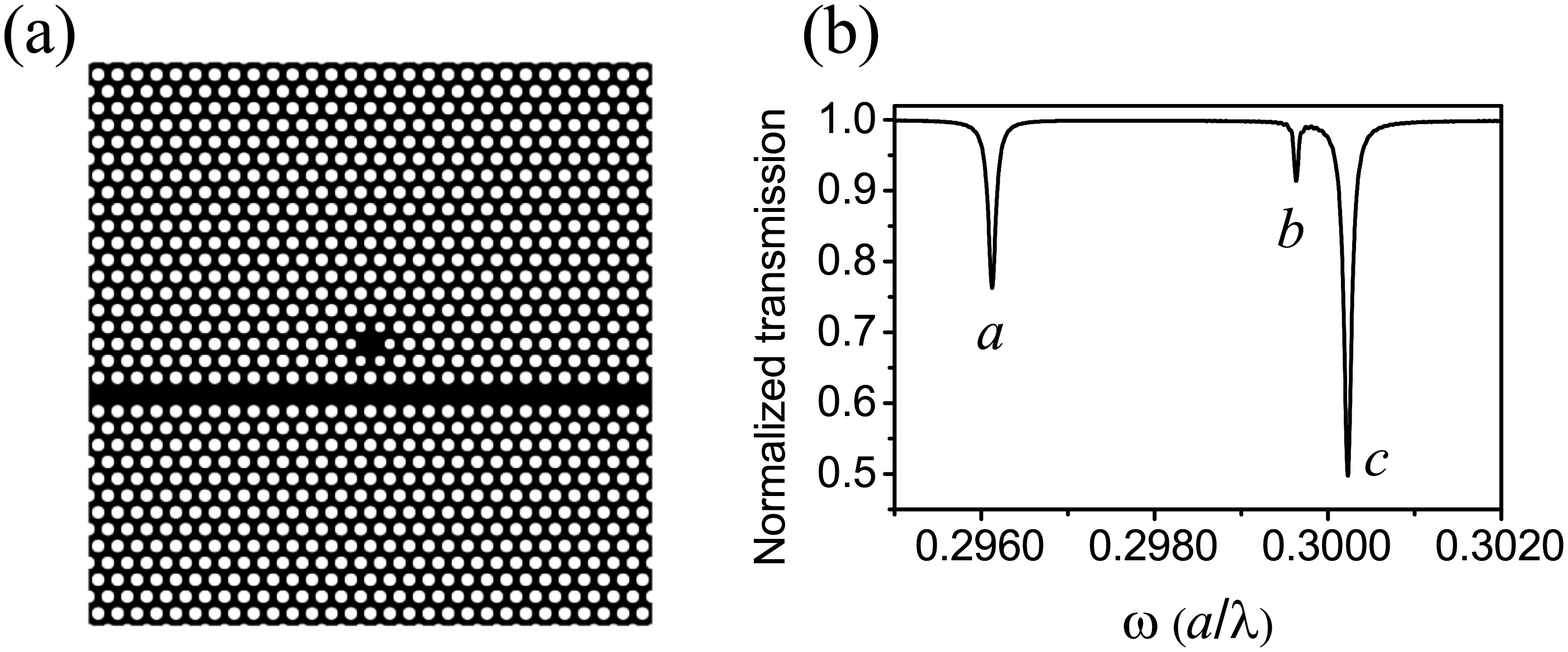}}
\caption{Coupled PhC H1 cavity-waveguide structure and transmission spectrum. (a) An H1 cavity is side-coupled with a W1 waveguide in a PhC slab with hexagonal lattice of air holes. The W1 waveguide is three lattice constant away from the center of the H1 cavity. (b) Transmission spectrum with light transmitting from one end of the waveguide, passing by the H1 cavity, and being detected at another end of the waveguide. Three dips in the transmission spectrum represent three cavity modes of the H1 cavity.} \label{scheme}
\end{figure}

\clearpage
\begin{figure}
\centering \resizebox{16cm}{!}{\includegraphics{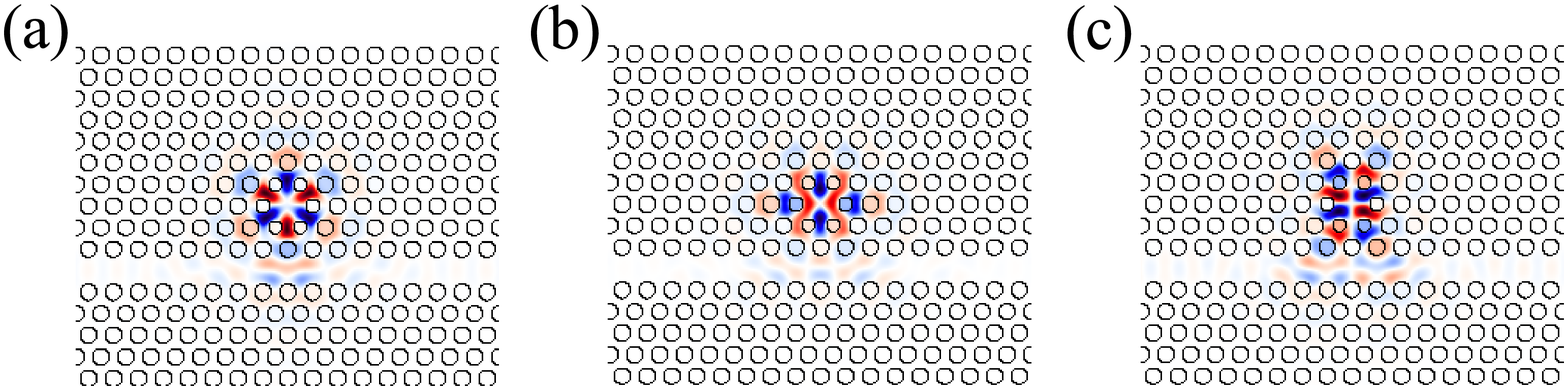}}
\caption{$\textit{H}_{z}$ field distributions of the cavity modes corresponding to the three dips which are denoted with \textit{a, b, c} in the transmission spectrum in \mbox{Fig. 1(b)}.} \label{g2g}
\end{figure}

\clearpage
\begin{figure}
\centering \resizebox{16.5cm}{!}{\includegraphics{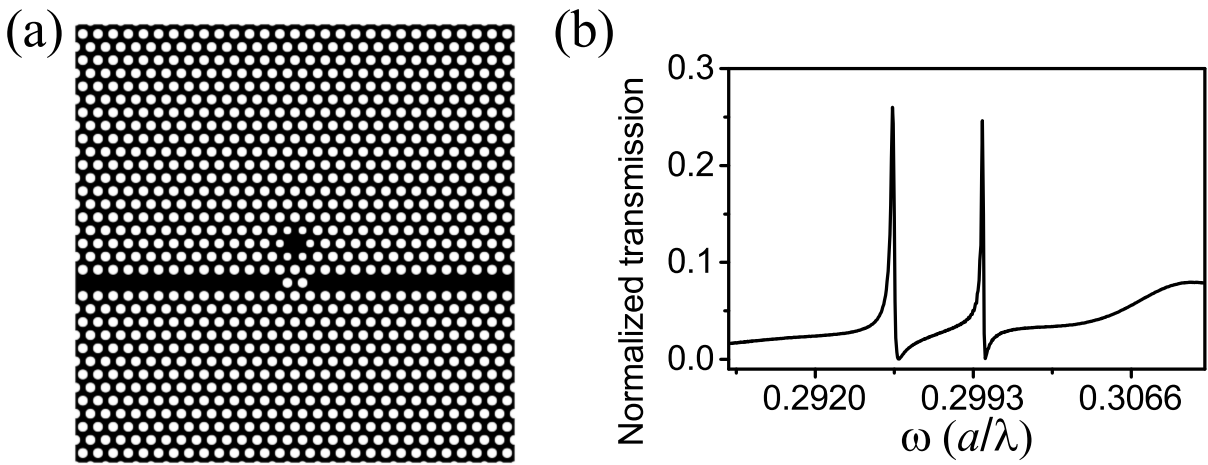}}
\caption{(a) Coupled cavity-waveguide structure with partially reflecting elements. (b) The transmission spectrum of the coupled structure. Two asymmetric Fano resonance line shapes appear due to the coupling between cavity and waveguide with partially reflecting elements.}
\end{figure}

\clearpage
\begin{figure}
\centering \resizebox{12cm}{!}{\includegraphics{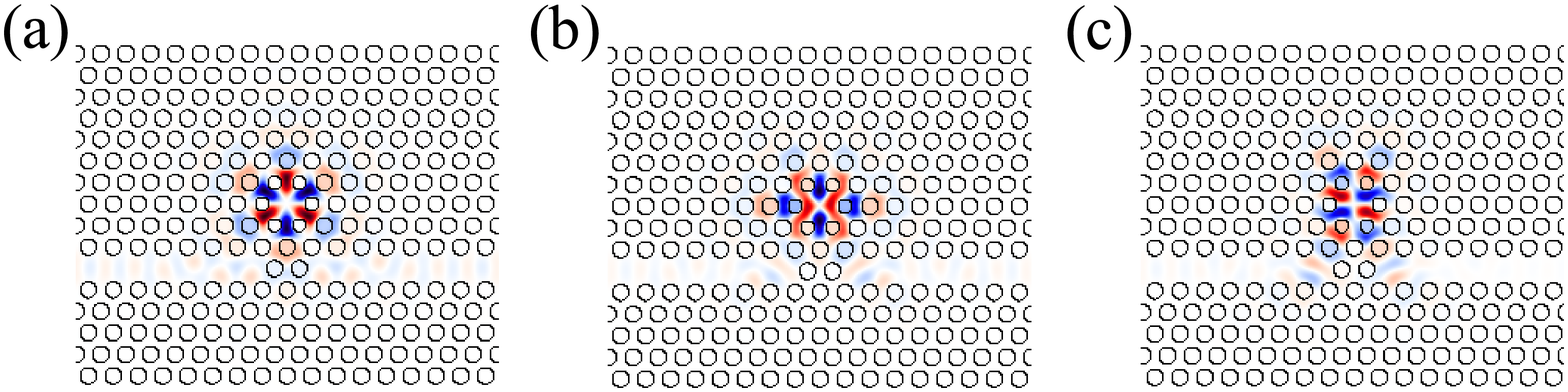}}
\caption{$\textit{H}_{z}$ field distributions of the cavity modes corresponding to the two Fano-type peaks in the transmission spectrum in \mbox{Fig. 3(b)}. (a) is the hexapole mode with a frequency of 0.2956 (\textit{a}/$ \lambda $). (b) and (c) are the degenerated quadrupole modes with frequency at 0.2998 (\textit{a}/$ \lambda $).}
\end{figure}

\clearpage
\begin{figure}
\centering \resizebox{16cm}{!}{\includegraphics{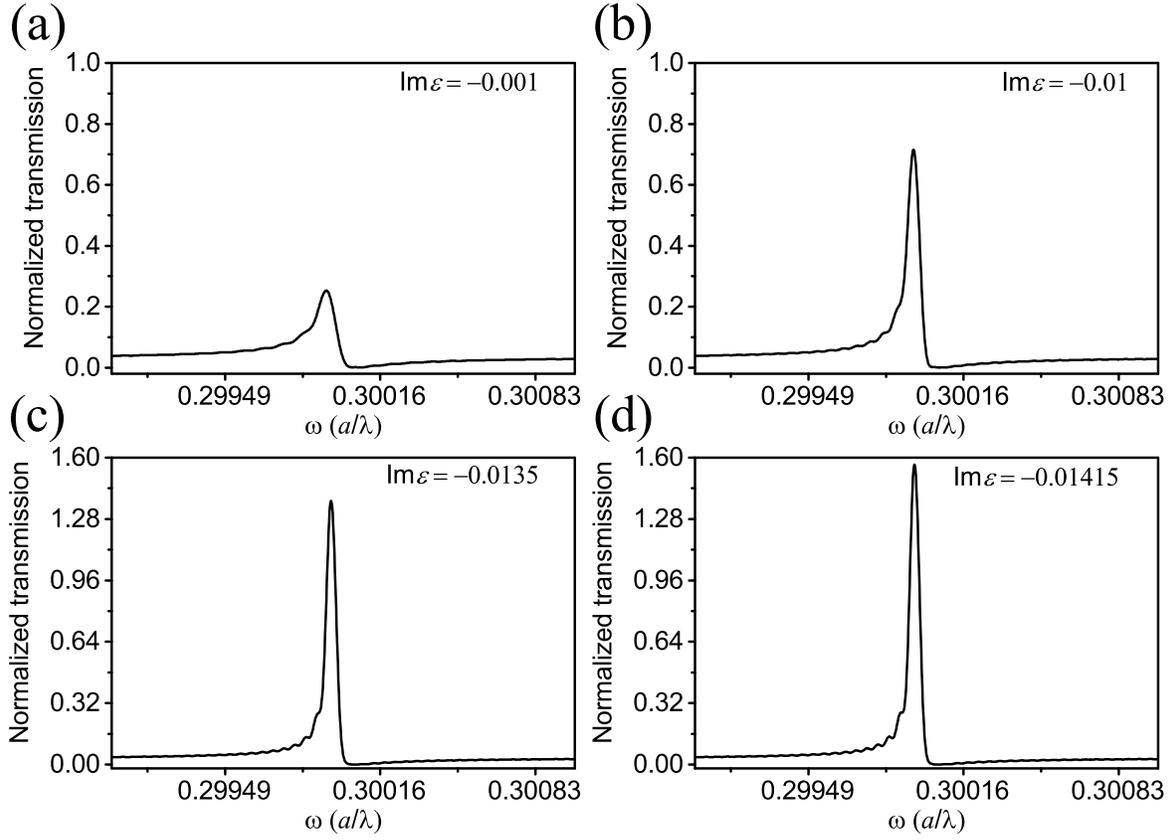}}
\caption{The transmission spectra of the coupled PhC cavity-waveguide structure with gain material introduced into the cavity. As the amount of gain increases from (a) to (d), the normalized transmissivity increases.} \label{Omega-Delta}
\end{figure}

\end{document}